\documentclass[12pt]{iopart}
\usepackage{iopams}  
\usepackage{theorem}

\theorembodyfont{\rmfamily}
\theoremstyle{plain}
\newtheorem{definition}{Definition}

\newcommand{\iv}{{}^4\!}
\newcommand{\tM}{\tilde{M}}  
\newcommand{\tg}{\tilde{g}} 
\newcommand{\Mg}[1]{(\iv M_{#1},g_{#1})}
\newcommand{\tMg}[1]{(\iv \tM_{#1},\tg_{#1})}
\renewcommand{\th}{\tilde{h}} 
\newcommand{\Mt}{M_{t}}
\newcommand{\Mh}[1]{(M_{#1 t},h_{#1})}
\newcommand{\tMh}[1]{(\tM_{#1 t},\th_{#1})}
\newcommand{\Mq}[1]{(\del \iv M_{#1},q_{#1})}
\newcommand{\Ss}[1]{(\Sigma_{#1},\sigma_{#1})}

\newcommand{\I}{{\rm I}}  

\newcommand{\VI}{{\rm VI}_0}
\newcommand{\const}{{\rm constant}}  
\newcommand{\z}{\zeta}
\newcommand{\Isom}[1]{{\rm Isom}(#1)}
\newcommand{\Esom}[1]{{\rm Esom}(#1)}

\newcommand{\columns}[2]%
{\left(\begin{array}{c} #1 \\ #2 \end{array}\right)}
\newcommand{\columnt}[3]%
{\left(\begin{array}{c} #1 \\ #2 \\ #3 \end{array}\right)}
\newcommand{\matrixs}[4]%
{\left( \begin{array}{cc}#1 & #2 \\ #3 & #4 \end{array} \right)}

\newcommand{\nab}{\nabla}
\newcommand{\del}{\partial}  
\newcommand{\vect}[1]{\frac{\del}{\del #1}}
\newcommand{\kvec}[1]{\Big( \frac{\del}{\del #1} \Big)}
\newcommand{\dd}[2]{\frac{d#1}{d#2}}
\newcommand{\kdd}[2]{\Big( \frac{d#1}{d#2} \Big)}

\newcommand{\bk}[2]{\langle #1 \ |\ #2 \rangle}
\newcommand{\inv}{^{-1}}
\newcommand{\sign}{{\rm sign}}

\begin{document}
\title{Thurston's Geometrization Conjecture and cosmological models}
\author{Katsuhito Yasuno\dag, Tatsuhiko Koike\ddag\ and Masaru Siino\dag}
\address{\dag\ Department of Physics, Tokyo Institute of Technology, 
Oh-Okayama, Meguro-ku, Tokyo 152-0033, Japan}
\address{\ddag\ Department of Physics, Keio University, 
Hiyoshi, Kohoku, Yokohama 223-8522, Japan}
\address{\ddag\ Department of Physics, University of Maryland, 
College Park, MD 20742-4111, USA}

\begin{abstract}
We investigate a class of spatially compact inhomogeneous spacetimes. 
Motivated by Thurston's Geometrization Conjecture, we give a formulation 
for constructing spatially compact composite spacetimes as solutions for 
the Einstein equations. 
Such composite spacetimes are built from the spatially compact locally 
homogeneous vacuum spacetimes which have two commuting Killing vectors 
by gluing them through a timelike hypersurface admitting a homogeneous 
spatial slice spanned by the commuting Killing vectors. 
Topology of the spatial section of the timelike boundary is taken to be 
the torus. 
We also assume that the matter which will arise from the gluing is 
compressed on the boundary, i.e. we take the thin-shell approximation. 
By solving the junction conditions, we can see dynamical behavior of 
the connected (composite) spacetime. 
The Teichm\"uller deformation of the torus also can be obtained. 
We apply our formalism to a concrete model. 
The relation to the torus sum of 3-manifolds and the difficulty of 
this problem are also discussed.  
\end{abstract}

%\pacs{xxx}

%\submitto{\CQG}
\maketitle

%%%%%%%%%%%%%%%%  Section 1
\section{Introduction}
\label{sec:Introduction}

A compact 3-manifold is an essential mathematical framework to study 
the evolution of the universe in the gravitational viewpoint. 
Nevertheless to prepare a concrete compact 3-manifold which is useful 
for investigating the universe we cannot help assuming very high symmetries. 
Especially, homogeneous compact 3-spaces are representative but 
as well-discussed it is very special situation. 
Then, how can we get more general compact 3-manifold and study it? 

The well-discussed direction is to reduce the symmetry of compact 3-manifolds. 
A remarkable trial is to introduce the Gowdy spacetime which admits 
the ${\rm U}(1)\times {\rm U}(1)$ symmetric metric with 
spacelike group orbits~\cite{GOWDY74}. Since it contains partial inhomogeneity 
(along one direction), some important researches have been done on it 
about the role of its inhomogeneity. Inherently, however, 
the Gowdy spacetime must have the same topology as corresponding 
compact homogeneous spacetime~\cite{TANIMOTO98}. 
So, it will not do for exploring the role of topology in general relativity 
(we know a unique exception~\cite{TANIMOTO00}). 

On the other hand, to grope for the way to approach compact 3-manifolds 
with general topologies, well believed Thurston's conjecture becomes 
a good guiding principle. This conjecture, which is called the 
\emph{Geometrization Conjecture}, states that 
any compact 3-manifold can be decomposed along embedded two-spheres and tori 
into domains, each of which admits one of eight types of geometric structure, 
i.e. a locally homogeneous metric~\cite{THURSTON82}. 
From this conjecture we expect that each compact 3-manifolds can be 
modeled on by an appropriate combination of homogeneous building blocks and 
the non-trivial connections between them. 
When we hope to deal with a compact 3-manifold with general topology 
in the general relativistic context, it is important to understand 
how the dynamical degrees of freedom are assigned to. 
In the viewpoint of the Geometrization Conjecture, we may say that 
the whole of the dynamical degrees of freedom of the compact composite 
3-manifold is substituted for by those of the compact homogeneous manifolds 
and the matter connecting them. 

Such a composite structure has never been regarded important in the context of 
general relativity. Of course, it might be natural to consider that 
our spacetime has no composite structure. 
Nevertheless we consider the composite structure possesses important meanings 
because to understand the relation between the topology and 
geometrical dynamics totally, we should prepare all 3-manifolds and 
study their geometrical dynamics. Furthermore, 
this total study about the topology and geometrical dynamics 
will become strong aim to explore the general evolution of all solutions of 
the Einstein equations and the universe created through quantum process. 

Before we describe our procedure to construct the composite spacetimes in detail, 
we clarify the concept of the composite structure of the 3-manifold. 
A compact 3-manifold is \emph{composite} if 
it admits non-trivial prime decomposition, i.e. the decomposition along 
the embedded two-spheres. 
The prime decomposition yields three kinds of prime manifolds, 
spherical types (which, if the elliptization conjecture is true, are 
compact quotients of the homotopy three-spheres by the discrete subgroups of 
${\rm SO}(4)$ acting freely and orthogonally on the homotopy three-spheres), 
$\mathbb{S}^2 \times \mathbb{S}^1$'s, and 
$K(\pi,1)$'s (which are irreducible and their universal covers are contractable). 
While the first two components cannot be decomposed any more, 
the third ones may subject to the torus decomposition~\cite{SCOTT83}. 
Then, finally one can get the most simple pieces of compact 3-manifolds, 
i.e. if the Geometrization Conjecture is true, they are modeled on one of the eight 
model geometries. 

Morrow-Jones and Witt~\cite{MJW93} studied about spacetimes with 
general topology by constructing the compact locally spherically symmetric 
3-manifolds glued along two-spheres, i.e. connected sum of 3-manifolds. 
However, the torus-sum of 3-manifolds has not been studied yet. 
Our study was initially motivated by the shortage of understanding of 
the latter kinds of 3-manifolds in general relativity. 

Our basic knowledge needed below is that of the spacetime constructed from 
a compact locally homogeneous 3-manifold. Fortunately, 
it is well understood in connection with extensive study of 
the Bianchi models~\cite{RS75}, i.e. spacetimes with \emph{global} spatial 
homogeneity which is implicitly assumed to be open except the Bianchi type IX 
that is naturally closed~\cite{AS91,FIK93,KTH94,TKH97a,TKH97b,KODAMA98}. 
The most important progress of these studies of spatially compact locally 
homogeneous spacetimes (SCLHSs) is that they succeeded to bring about 
\emph{global} degrees of freedom associated with compactifications of 
spatial sections, which, in fact, carry the \emph{dynamical} degrees of freedom. 
More precisely, applying a prescription given by Tanimoto, Koike and Hosoya 
(TKH)~\cite{TKH97a}, we can consider non-isometric deformations of 
compact locally homogeneous spatial hypersurfaces leaving 
their universal cover globally conformally isometric, so called 
the \emph{Teichm\"uller deformation}. 
(Because their approach is crucial for our purpose of this paper,  
we briefly review their construction of SCLHSs in section~\ref{sec:TKH}.) 

Our strategy for constructing spatially compact composite spacetimes is 
the direct application of TKH's construction of SCLHSs. 
Briefly, the SCLHSs are regarded as the building blocks of composite spacetimes. 
As mentioned above, other dynamical degrees of freedom are assigned to 
the matter connecting them.  
As a simple setup, we assume that the dynamics of the matter is 
approximated by a thin shell. That is, 
the well-known junction method~\cite{ISRAEL66} is applied to connect the SCLHSs. 
Of course, this simplification might discard several important features 
of inhomogeneity. Nevertheless we believe that we can recognize 
some interesting aspects of the compact manifold with general topology. 

Let us assume that two SCLHSs, each of which contains 
a \emph{commuting} pair of the Killing vectors and satisfies 
the Einstein vacuum equations, are given. 
The existence of the commuting Killing vectors ensures the SCLHS has 
a \emph{homogeneous} torus section since 
the fundamental group of the spatial section contains at least one 
commuting relation. 
Then cut each SCLHS along a timelike hypersurface and glue the resulting boundaries 
which admits a foliation by spatial leaves diffeomorphic to the homogeneous tori 
so as to satisfy the junction conditions, and one get a composite spacetime as 
a solution of the Einstein equations. 

The organization of this paper is as follows. 
In section~\ref{sec:TKH}, we briefly review the construction of SCLHSs 
along the line of~\cite{TKH97a}. 
In section~\ref{sec:formalism}, we formulate the way of constructing 
the spatially compact composite spacetimes with the help of 
the prescription given in section~\ref{sec:TKH} and 
the well-known junction method of Israel~\cite{ISRAEL66}. 
In section~\ref{sec:example}, we apply the formalism given in 
the preceding section to construct a simple model, 
a composite spacetime whose spatial section consists of 
a compact quotient modeled on $\mathbb{E}^3$ with 
a vacuum Bianchi type I solution (Kasner solution) of the Einstein equations and 
a compact quotient modeled on Sol with a vacuum Bianchi type VI$_0$ solution 
(Ellis-MacCallum solution~\cite{EM69}) of the Einstein equations. 
We shall see there the spacelike section of the glued spacetime is 
topologically a compact quotient of Sol. Also, the dynamical behavior of 
the resulting spacetime is discussed. 
Summary and discussions is asserted in section~\ref{sec:smry_dcs}. 

%%%%%%%%%%%%%%%%  Section 2
\section{Construction of SCLHSs}\label{sec:TKH}

Let us begin by setting SCLHSs as building blocks of compact composite spacetimes. 
The construction of SCLHSs have been given by 
Tanimoto \etal~\cite{TKH97a}. We briefly review their construction. 

Let $\Mg{}$ be a smooth Lorentzian four-dimensional spacetime. 
We assume that $\Mg{}$ admits a foliation by compact locally 
homogeneous spatial leaves $\Mh{}$, where $t = \const$ labels 
each leaf of the foliation and $h$ is the induced three-dimensional 
Riemannian metric on $\Mt$. A metric on a manifold $M$ is 
\emph{locally homogeneous} if for any points $x,y \in M$ there 
exist neighborhoods $U,V \in M$ and an isometry $(U,x) \to (V,y)$. 
Such a spacetime $\Mg{}$ is called a spatially compact locally 
homogeneous spacetime (SCLHS). We denote the universal cover 
of $\Mg{}$ as $\tMg{}$. $\Mg{}$ is obtained from $\tMg{}$ by 
identifying the spatial leaf $\tMh{}$ using a discrete subgroup 
$\Gamma$ of an \emph{extendible isometry group}, a subgroup of 
an isometry of $\tM_t$ which preserves $\tM_t$, acting freely 
on $\tM_t$. 

The reason why we consider not the isometry but the extendible isometry is 
that if we compactify with $\Gamma \subset \Isom{\tM_t}$, 
resulting quotient $\Mg{}$ will not, in general, be remained smooth~\cite{TKH97a}.  
The definition of the extendible isometry group is given as follows~\cite{TKH97a}: 

\begin{definition}
Let $\tMh{}$ be a spatial leaf of $\tMg{}$. 
An \emph{extendible isometry} is the restriction on $\tM_t$ of an isometry 
of $\tMg{}$ which preserves $\tM_t$. 
They form a subgroup of $\Isom{\tM_t}$. 
We call it the \emph{extendible isometry group}, and denote it as 
$\Esom{\tM_t}$. 
Obviously, an extendible isometry $a \in \Esom{\tM_t}$ has 
the natural extension on $\iv \tM$ which is an element of $\Isom{\iv \tM}$ 
and preserves $\tM_t$. We call such the natural extension on $\iv \tM$ 
the extended isometry of $a$, or simply the extension of $a$. 
\end{definition}

From this definition, we can conclude that the identifications on 
the initial surface $\tMh{}$ must be implemented in $\Esom{\tM_t}$, 
$\Gamma \subset \Esom{\tM_t}$, where $\Gamma$ is a discrete subgroup of 
$\Esom{\tM_t}$ acting freely on $\tM_t$, 
to get a SCLHS out of a given four-dimensional universal cover $\tMg{}$. 
Moreover, the identifications acting on whole $\tMg{}$ are determined by 
the action of the extension of $\Gamma$ on $\iv \tM$. 
We should remark that all the well-known Bianchi groups are, 
by definition, the extendible isometry groups. 

Global deformations of SCLHSs are expressed by the evolution 
of \emph{Teichm\"uller  parameters} which spans the 
\emph{Teichm\"uller space}, $\mathcal{T}(M)$, where $M$ is a compact 3-manifold. 
We here take the definition of the Teichm\"uller space same as in 
Koike, Tanimoto and Hosoya~\cite{KTH94}. 

\begin{definition}
Let $M$ be a compact connected orientable manifold of dimension 
2 or 3, and $\tM$ be its universal cover. 
Let ${\rm Rep}(M)$ be the space of all discrete and faithful representations 
$\rho : \pi_1(M) \to \Isom{\tM}$, where $\pi_1(M)$ is the fundamental group of $M$. 
In other words, ${\rm Rep}(M)$ is the space of all covering groups $\Gamma$ of $\tM$. 
A diffeomorphism $\phi : \tM \to \tM$ such that 
$\phi_* \th_{ab} = f \th_{ab}$, where $f$ is a constant, is called 
a globally conformal isometry. 
Let ${\rm GCI}(\tM)$ be the space of all such diffeomorphisms. 
The equivalence relation in ${\rm Rep}(M)$ is defined by the conjugation such that: 
for $\rho, \rho' \in {\rm Rep}(M)$ and $\alpha \in \pi_1(M)$, 
\begin{equation}
\rho'(\alpha) = \phi \circ \rho(\alpha) \circ \phi\inv .
\end{equation}
Then the \emph{Teichm\"uller space} is defined by the orbit space such that 
\begin{equation}
\mathcal{T} (M) = \frac{{\rm Rep} (M)}{{\rm GCI} (\tM)} .
\end{equation}
\end{definition}

Let us give the definition of the Teichm\"uller deformation~\cite{TKH97a}. 

\begin{definition}
The \emph{Teichm\"uller deformation} is the smooth, non-isometric 
deformation of spatial metric $h$ on $M_t$ which leaves the 
universal cover $\tMh{}$ globally conformally isometric. 
\end{definition}

We shall study how we can obtain such deformations. Consider 
an universal cover $\tMg{}$. A form of the metric $\tg$ can be 
represented by taking the synchronous gauge as~\cite{RS75},
\begin{equation}
\label{def:synchro}
\tg_{ab} 
= -(dt)_a (dt)_b + \th_{ij}(t) (\chi^i)_a (\chi^j)_b ,
\end{equation}
where $(\chi^i)_a$ is the invariant dual basis. 
Note that for vacuum solutions of the Einstein equations for 
Bianchi class A~\cite{EM69} (i.e. types I, II, VI$_0$, VII$_0$, 
VIII, and IX) and type V, the metric components $\th_{ij}(t)$ in \eref{def:synchro} 
become diagonal by diffeomorphisms~\cite{SK94} such that 
\begin{equation}
\label{eq:synchro_diag}
\fl 
\tg_{ab} 
= -(dt)_a (dt)_b + \th_{11}(t) (\chi^1)_a (\chi^1)_b 
+ \th_{22}(t) (\chi^2)_a (\chi^2)_b + \th_{33}(t) (\chi^3)_a (\chi^3)_b . 
\end{equation}

Next consider the initial identification, 
$\tM_t \to \tM_t/\Gamma = M_t$, where $\Gamma \in \Esom{\tM_t}$. 
By taking conjugations by the global conformal isometry with 
respect to $\Esom{\tM_t}$, we obtain a representation up to 
conjugation such that: for $\alpha,\beta,\gamma \in \pi_1(M_t)$, 
\begin{equation}
\rho : \{ \alpha,\beta,\gamma \} \mapsto 
\Gamma 
= \left\{ \columnt{a^1}{a^2}{a^3}, \columnt{b^1}{b^2}{b^3}, 
\columnt{c^1}{c^2}{c^3} \right\} ,
\end{equation}
where each component vector forms $\Esom{\tM_t}$, i.e., 
\begin{equation}
\left\{ \columnt{a^1}{a^2}{a^3} 
:= e^{a^3 \xi_{3}}e^{a^2 \xi_{2}}e^{a^1 \xi_{1}} \ |\ a^i \in \mathbb{R} \right\}, 
\end{equation}
where $\xi_i$'s are the (local) Killing vectors which form the Lie 
algebra of $\Esom{\tM_t}$. 

Finally, we should consider how to relate the initial identification $\Gamma$ 
with given Teichm\"uller parameters which are defined on 
a compact 3-manifold intrinsically. 
By noting that the local structures of $\Mh{}$ are naturally inherited from 
its universal cover $\tMh{}$, the induced metric $h$ on $M_t$ is given by
\begin{equation}
\label{eq:ind_metric_diag}
h_{ab} 
= h_{11}(t) (\chi^1)_a (\chi^1)_b + h_{22}(t) (\chi^2)_a (\chi^2)_b 
+ h_{33}(t) (\chi^3)_a (\chi^3)_b .
\end{equation}
The functions $h_{ii}(t)$'s are constant on $M_t$. 
Hence, we can transform this metric by using the 
\emph{homogeneity preserving diffeomorphism}~\cite{AS91,KTH94} to 
\begin{equation}
\label{eq:h_afterHPD}
h_{ab} = f \delta_{ij} (\chi^{\prime i})_a (\chi^{\prime j})_b ,
\end{equation}
where $f$ is a constant on $M_t$, $(\chi^{\prime i})_a$ is an invariant 
dual basis expressed in terms of the transformed coordinate. 
By this transformation, $\Gamma$ is also transformed and we denote it 
with a prime, $\Gamma'$. In fact, $\Gamma'$ includes the metric components, 
i.e. $\Gamma' = \Gamma(t)$. 
Since the metric (\ref{eq:h_afterHPD}) is conformally equivalent to 
the standard metric on $M_t$, 
$h^{{\rm standard}}_{ab} = \delta_{ij} (\chi^i)_a (\chi^j)_b$, 
we can take conjugations of $\Gamma'$ by the global conformal isometry 
with respect to $\Isom{\tM_t}$ so that we can relate the initial identification 
$\Gamma$ with the Teichm\"uller parameters given intrinsically on 
a compact 3-manifold $M$, $i(M) = M_t$, where $i$ denotes the embedding of $M$ 
into $\iv M$. Thus, by the fact that $\Gamma'$ involves the metric components, 
the Teichm\"uller parameters can be expressed by the metric components 
appearing in \eref{eq:ind_metric_diag} and the parameters of initial identification. 
In this way, we can regard the Teichm\"uller parameters as 
the dynamical variables. This finishes the task. 

%%%%%%%%%%%%%%%%  Section 3
\section{Formalism of the construction of composite spacetimes}
\label{sec:formalism}

In this section, we give a formulation to construct a composite spacetime 
which consists of SCLHSs as building blocks and timelike shells connecting them. 
As mentioned, we consider only the case that the timelike shells are foliated 
by homogeneous spacelike tori. 

As a fundamental construction, we consider the gluing of two \emph{different} 
SCLHSs along two timelike shells. (However, these two shells are equivalent.) 
Of course, we can apply the prescription given below to the gluings of many SCLHSs. 

%%%%%%%%%%%%%%%%  Subsection 3.1
\subsection{Topological decomposition}

Let $\Mg{A}$ and $\Mg{B}$ be two vacuum SCLHSs, each of which 
admits at least one pair of two \emph{commuting} local Killing vectors. 
That is, if $(\xi_A)_i$ and $(\xi_B)_i$ denote linearly independent 
local Killing vectors of SCLHSs, two of them in each SCLHS satisfies
\begin{equation}
[(\xi_A)_1,(\xi_A)_2] = 0, \quad 
[(\xi_B)_1,(\xi_B)_2] = 0 .
\end{equation}
The existence of the two commuting Killing vectors is the necessary condition 
so that we can cut out a locally homogeneous torus from 
a compact locally homogeneous 3-manifold. 
The fundamental groups $\pi_1(M_A)$ and $\pi_1(M_B)$ is also necessary to 
contain at least one commuting relation such that:
\begin{eqnarray}
\pi_1(M_A) 
&= \bk{ \alpha_A, \beta_A,\gamma_A }
        {[\alpha_A,\beta_A] = 1,\ {\rm the\ other\ two\ relations}} ,\\
\pi_1(M_B) 
&= \bk{ \alpha_B, \beta_B,\gamma_B }
        {[\alpha_B,\beta_B] = 1,\ {\rm the\ other\ two\ relations}} ,
\end{eqnarray}
where $[\alpha, \beta] := \alpha \beta \alpha\inv \beta\inv$. 

The first step is to cut out each SCLHS along a timelike hypersurface 
which admits a foliation by locally homogeneous tori. 
We denote two SCLHSs as $\Mg{X}$. Let $\Mq{X}$ be a timelike hypersurface to which 
two commuting Killing vectors are tangent. 
Let $\Ss{X}$ be a spatial section of $\Mq{X}$. 
Then each $\Ss{X}$ is the Killing orbits of two commuting Killing vectors 
$(\xi_X)_1$ and $(\xi_X)_2$. Therefore, $\Sigma_X$ is diffeomorphic to $T^2$ and 
is represented by the fundamental group,
\begin{equation}
\pi_1(\Sigma_X) = \bk{\alpha_X,\beta_X}{[\alpha_X,\beta_X] = 1} .
\end{equation}
Its representation up to conjugations of the globally conformal isometry 
with respect to $\Esom{\Sigma_X}$ is given by
\begin{equation}
\Gamma_X |_{\Sigma_X} 
= \{ \rho_X(\alpha_X),\rho_X(\beta_X) \} =: \{ a_X,b_X \}  
= \left\{ \columnt{a_X^1}{a_X^2}{a_X^3}, \columnt{b_X^1}{b_X^2}{b_X^3} \right\} .
\end{equation}
Obviously, however, this representation is redundant. 
Since only $(\xi_X)_1$ and $(\xi_X)_2$ lie in $\Sigma_X$, 
the third components $a_X^3$ and $b_X^3$ of the above representations, 
i.e. the components of the basis $(\xi_X)_3$, must be zero. 
Thus, the representation of $\pi_1(\Sigma_X)$ up to conjugations is given by
\begin{equation}
\Gamma_X |_{\Sigma_X} 
= \{ a_X|_{\Sigma_X},b_X|_{\Sigma_X} \} 
= \left\{ \columns{a_X^1}{a_X^2}, \columns{a_X^1}{b_X^2} \right\} .
\end{equation}
Also, the initial identification of $\tM_t$ is represented by
\begin{eqnarray}
\Gamma_X 
&= \{ \rho_X(\alpha_X),\rho_X(\beta_X), \rho_X(\gamma_X)\} 
=: \{ a_X,b_X,c_X \}  \nonumber\\
&= \left\{ \columnt{a_X^1}{a_X^2}{0}, \columnt{b_X^1}{b_X^2}{0}, 
\columnt{c_X^1}{c_X^2}{c_X^3} \right\} .
\end{eqnarray}

%%%%%%%%%%%%%%%%  Subsection 3.2
\subsection{Geometrical decomposition}

Consider the local (i.e. geometrical) description of $\Mq{X}$. 
Let us introduce the orthonormal basis on $\Mq{X}$, $\{ (\tau_X)^a, (e_{Xp})^a \}$ 
($p = 1,2$), where $(\tau_X)^a = (e_{X0})^a$ represents the timelike basis. 
By using its dual, $\{ (\tau_X)_a, (\theta_X^p)_a \}$, we can write the metric $q_X$ as 
\begin{equation}
(q_X)_{ab} 
= - (\tau_X)_a (\tau_X)_b + \delta_{pq}(\theta_X^p)_a (\theta_X^q)_b .
\end{equation}
Let $(n_X)^a$ be a unit normal to $\del \iv M_X$. 
It is convenient to introduce the \emph{Gaussian normal coordinate system} in 
the neighborhood of $\Mq{X}$ for expressing the embedding of $\Mq{X}$ into 
$\Mg{X}$~\cite{WALD84}. Let $\z_X$ be the Gaussian normal coordinate 
satisfying $\z_X = \const$ on $\Mq{X}$. 
For definiteness, we set $\z_X = 0$ on $\Mq{X}$. 
An orientation of the coordinate $\z_X$ will be defined when 
we perform the gluing. Then, in this neighborhood,  
we can write the four-dimensional metric $g_X$ as 
\begin{equation}
(g_X)_{ab} 
= (q_X)_{ab} + (n_X)_a (n_X)_b 
= (q_X)_{ab} + (d\z_X)_a (d\z_X)_b .
\end{equation}
The extrinsic curvature on $\del \iv M_X$ is defined by 
\begin{equation}
\label{eq:def_extrinsic_curv}
(K_X)_{ab} := (q_X)_a{}^m (q_X)_b{}^n \nab_m (n_X)_n ,
\end{equation}
where $(q_X)_a{}^m = \delta_a{}^m - (n_X)_a (n_X)^m$ is the projection operator 
onto $\del \iv M_X$, and $\nab_a$ is the derivative operator compatible with 
the spacetime metric $g_X$. 

%%%%%%%%%%%%%%%%  Subsection 3.3
\subsection{Gluing}

The next step is to carry out the gluing of two SCLHSs, $\Mq{A}$ and $\Mq{B}$. 
The gluing of the spacetime manifolds, 
$\del \iv M_A = \del \iv M_B =: \del \iv M$, requires 
\begin{equation}
\Sigma_A = \Sigma_B =: \Sigma .
\end{equation}
This gluing can be expressed as
\begin{equation}
\label{eq:glue_Sigma}
\Gamma_A |_{\Sigma_A} = \Gamma_B |_{\Sigma_B} =: \Gamma |_{\Sigma} .
\end{equation}
Note that it is not uniquely determined. 
Indeed, there is a degree of freedom of one parameter $\phi \in \mathbb{R}$ such that
\begin{equation}
\columns{\xi_{A1}}{\xi_{A2}} 
= \mathcal{R}(\phi) \columns{\xi_{B1}}{\xi_{B2}} ,
\end{equation}
where $\mathcal{R}(\phi)$ is the rotation matrix by angle $\phi$. 
Then defining a new set of commuting Killing vectors of $\Mg{B}$, 
\begin{equation}
\xi'_{B1} := \xi_{B1} \cos \phi - \xi_{B2} \sin \phi ,\quad 
\xi'_{B2} := \xi_{B1} \sin \phi + \xi_{B2} \cos \phi ,
\end{equation}
we can explicitly represent the topological gluing \eref{eq:glue_Sigma} such that
\begin{equation}
\label{eq:glue_Gamma}
\columns{a_A^1}{a_A^2} 
= \columns{a_B^1}{a_B^2} =: \columns{a^1}{a^2} ,\quad 
\columns{b_A^1}{b_A^2} 
= \columns{b_B^1}{b_B^2} =: \columns{b^1}{b^2} .
\end{equation}
Thus, we obtain a representation of the initial identification on $\Sigma$, 
\begin{equation}
\label{eq:initial_gluing}
\Gamma |_{\Sigma} 
= \left\{ \columns{a^1}{a^2}, \columns{b^1}{b^2} \right\} .
\end{equation}

When we glue $\del \iv M_A$ and $\del \iv M_B$ together, 
we require that the induced metrics on $\del \iv M_A$ and $\del \iv M_B$ are 
to be isometric, $(q_A)_{ab} = (q_B)_{ab} =: q_{ab}$. 
Using the orthonormal basis introduced above, we can express the condition such that
\begin{equation}
\label{eq:metric_conti}
(\tau_A)_a = (\tau_B)_a ,\quad (\theta^p_A)_a = (\theta^p_B)_a .
\end{equation}
However, the extrinsic curvatures on the timelike boundaries, in general, 
cannot be continuous but have discontinuity expressed by 
the surface energy-momentum tensor on $\del \iv M$. 
It is defined by the integral of the four-dimensional energy momentum tensor 
projected onto $\del \iv M$ over the infinitesimal interval 
$I=[-\epsilon,\epsilon]$ along the Gaussian normal coordinate orthogonal to 
$\del \iv M$ 
\begin{equation}
S_{ab} := \int_{-\epsilon}^{\epsilon} \iv T_{mn}q^m{}_a q^n{}_b d\zeta .
\end{equation}
That is, we consider the case such that there is a delta-function singularity 
on $\del \iv M$. We define the region of $\z > 0$ to be $\Mg{A}$ and 
that of $\z < 0$ to be $\Mg{B}$. 

Thus, the \emph{junction condition}, which was formulated 
by Israel~\cite{ISRAEL66}, is given by
\numparts 
\begin{eqnarray}
\label{eq:H_constraint}
S^{mn}\{ K_{mn} \} &= [ \iv T_{mn} n^m n^n ] ,\\
\label{eq:M_constraint}
D^m S_{ma} &= - [ \iv T_{mn} n^m q^n{}_a] , \\
\label{eq:evolution}
[ K_{ab} ] &= - 8\pi \left( S_{ab} - \frac12 S q_{ab} \right) ,
\end{eqnarray}
\endnumparts
where $D_a$ is the derivative operator compatible with the induced metric 
$q_{ab}$ on $\del \iv M$, $S = q^{mn} S_{mn}$, and 
we have used the following notations,
\begin{eqnarray*}
[ \Psi ] 
&:= \lim_{\epsilon \to 0} 
        ( \Psi_A |_{\z = \epsilon} - \Psi_B |_{\z = -\epsilon} ) ,\\
\{ \Psi \} 
&:= \frac12 \lim_{\epsilon \to 0} 
        ( \Psi_A |_{\z = \epsilon} + \Psi_B|_{\z = -\epsilon} ) .
\end{eqnarray*}
By solving the above junction condition, we can obtain the dynamics of $\del \iv M$, 
i.e. the metric on $\del \iv M$ is determined. 

To take this opportunity, we remark that the meaning of applying 
the thin-shell approximation. 
Since we are trying to glue two SCLHSs admitting different geometric structures, 
to glue them smoothly we need to put a finite transition region with 
(at least one-dimensional) inhomogeneity between them. 
However, such a prescription forces us to analyze 
infinite dimensional dynamical degrees of freedom and 
investigate the existence of a solution of the Einstein equations 
for the system with which we are concerned. 
Unfortunately, we have not succeeded to demonstrate it. 
So, we adopt the thin-shell approximation to investigate the dynamics of 
the composite universe explicitly. 

Finally, we can consider the Teichm\"uller deformation of $\Sigma = T^2$. 
We recall that the Teichm\"uller space of the torus, $\mathcal{T}(T^2)$, is 
homeomorphic to $\mathbb{R}^2$~\cite{THURSTON97} and 
so it is represented, up to conjugations, 
by two Teichm\"uller parameters, $r, s \in \mathbb{R}$, $r > 0$, such that 
\begin{equation}
\label{eq:Teich_torus}
\Lambda = \left\{ \columns{r}{0} ,\columns{s}{1/r} \right\} .
\end{equation}
Then with the solution obtained by solving the junction condition, 
we can relate the initial identification \eref{eq:initial_gluing} 
with the Teichm\"uller parameters \eref{eq:Teich_torus} by applying 
the prescription given in section~\ref{sec:TKH}, so we can obtain 
the Teichm\"uller deformation of the boundary torus. 

In the next section, we shall apply this formalism to a concrete example. 
We consider the gluing of compact quotients of 
vacuum Bianchi type I and VI$_0$ universes for which the exact solutions 
have been known. 

%%%%%%%%%%%%%%%%  Section 4
\section{An Example}
\label{sec:example}

We study here an example of the composite spacetimes. 
It consists of compact quotients of vacuum Bianchi I and VI$_0$ 
spacetimes whose corresponding model geometries are $\mathbb{E}^3$ and Sol, 
respectively. In subsection~\ref{subsec:building_blocks}, 
we prepare the two SCLHSs and cut them along timelike hypersurfaces 
which admit a foliation by locally homogeneous tori. 
In subsection~\ref{subsec:dynamics}, we perform the gluing. 

%%%%%%%%%%%%%%%%  Subsection 4.1
\subsection{Preparation of SCLHSs}
\label{subsec:building_blocks}

%%%%%%%%%%%%%%%%  Subsubsection 4.1.1
\subsubsection{Compact quotient of vacuum Bianchi type I spacetime}

Bianchi I group is the 3-dimensional translation group $\mathbb{R}^3$ and 
characterized by the Lie algebra
\begin{equation}
[ (\xi_{\I})_i , (\xi_{\I})_j ] = 0 \quad (i,j = 1,2,3) ,
\end{equation}
for three Killing vectors $(\xi_{\I})_i$. 
In terms of the coordinate basis $(\xi_{\I})_i$'s are given by
\begin{equation}
(\xi_{\I})_1 = \vect{x_{\I}} ,\quad 
(\xi_{\I})_2 = \vect{y_{\I}} ,\quad 
(\xi_{\I})_3 = \vect{z_{\I}} ,
\end{equation}
where $(x_{\I},y_{\I},z_{\I}) = (x^1,x^2,x^3)$. 
The finite actions generated by these $(\xi_{\I})_i$'s are given by
\begin{equation}\label{eq:multiplication_I}
\columnt{a}{b}{c} \columnt{x}{y}{z} = \columnt{a+x}{b+y}{c+z} ,
\end{equation}
where the component vectors 
\begin{equation}
G_{\I} 
= \left\{ \columnt{a}{b}{c} 
:= e^{c (\xi_{\I})_3} e^{b (\xi_{\I})_2} e^{a (\xi_{\I})_1} \ |\ a,b,c \in \mathbb{R} \right\}
\end{equation}
form the Bianchi I group. 
The invariant dual basis of Bianchi I group is given by 
\begin{equation}
(\chi_{\I})^1 = dx_{\I} ,\quad 
(\chi_{\I})^2 = dy_{\I} ,\quad 
(\chi_{\I})^3 = dz_{\I} .
\end{equation}

Then the invariant metric on $\Mh{\I}$ inherited from its universal cover 
$\tMh{\I}$ can be written as 
\begin{equation}
\label{eq:spatialmetric_I}
(h_{\I})_{ab} = (h_{\I})_{ij} (\chi_{\I}^i)_a (\chi_{\I}^j)_b ,
\end{equation}
where $(h_{\I})_{ij}$ is a constant matrix. 

Let $\tMg{\I}$ be a vacuum, spatially homogeneous spacetime which has 
the Bianchi I group acting transitively on $\tM_{\I}$. 
The general Bianchi I vacuum solution of the Einstein equations is 
well known as the Kasner spacetime given by
\begin{equation}
\label{eq:Kasner_sol}
\fl 
(\tg_{\I})_{ab} 
= -(dt_{\I})_a (dt_{\I})_b 
+ t_{\I}^{2p_1} (dx_{\I})_a (dx_{\I})_b 
+ t_{\I}^{2p_2} (dy_{\I})_a (dy_{\I})_b 
+ t_{\I}^{2p_3} (dz_{\I})_a (dz_{\I})_b , 
\end{equation}
where $\sum_{i=1}^3 p_i = \sum_{i=1}^3 p_i^2 = 1$. 
The space of solutions (modulo isometries) is characterized by just one 
parameter which takes on a circle in $(p_1,p_2,p_3) \in \mathbb{R}^3$, 
the intersection of the unit 2-sphere with the plane. 

Now we consider the initial identification of $\tMg{\I}$ which yields a SCLHS, 
$\Mg{\I}$. For a compact quotient of $\tMh{\I}$, we take a three-torus, $T^3$, 
modeled on $\mathbb{E}^3$, whose fundamental group is given by
\begin{equation}
\pi_1(M_{\I}) 
= \bk{ \alpha_{\I},\beta_{\I},\gamma_{\I} }
{ [\alpha,\beta] = 1,[\beta,\gamma]=1,[\gamma,\alpha]=1 }.
\end{equation} 
Its representation up to conjugations with respect to the Bianchi I group 
(however, no non-trivial conjugation exists, see \eref{eq:multiplication_I}) 
is given by
\begin{equation}
\Gamma_{\I} 
= \{ a_{\I},b_{\I},c_{\I} \} 
= \left\{ \columnt{a_{\I}^1}{a_{\I}^2}{a_{\I}^3} , 
\columnt{b_{\I}^1}{b_{\I}^2}{b_{\I}^3} , \columnt{c_{\I}^1}{c_{\I}^2}{c_{\I}^3} \right\} .
\end{equation}
These three vectors must be linearly independent. 

We cut out the spacetime $\Mg{\I}$ along a timelike hypersurface $\Mq{\I}$ 
whose spatial section, $\Ss{\I}$, is taken to be the Killing orbits of $(\xi_{\I})_1$ 
and $(\xi_{\I})_2$. It means that if we choose the generators 
$\alpha_{\I}$ and $\beta_{\I}$ as the generators of the boundary torus 
$\Sigma_{\I} =T^2$, their representations must satisfy $a_{\I}^3 = 0 = b_{\I}^3$. 
Moreover, in this case, the remaining vector $c_{\I}$ must satisfy 
$c_{\I}^3 \neq 0$ so that the three vectors $a_{\I}$, $b_{\I}$ and $c_{\I}$ 
are to be linearly independent. 
Thus, we obtain the representation on $\Sigma_{\I}$,
\begin{equation}
\label{eq:initial_I}
\Gamma_{\I}|_{\Sigma_{\I}} 
= \left\{\columns{a_{\I}^1}{a_{\I}^2} ,\columns{b_{\I}^1}{b_{\I}^2} \right\} .
\end{equation}

Next we consider the local geometry of $\Mq{\I}$. 
Since $(\xi_{\I})_1 = \del/\del x_{\I}$ and 
$(\xi_{\I})_2 = \del/\del y_{\I}$ lie in $\del \iv M_{\I}$, 
the unit future-directed timelike vector $(\tau_{\I})^a$ can be written as
\begin{equation}
(\tau_{\I})^a 
= \kvec{\tau_{\I}}^a 
= \dd{t_{\I}}{\tau_{\I}} \kvec{t_{\I}}^a + \dd{z_{\I}}{\tau_{\I}} \kvec{z_{\I}}^a .
\end{equation}
From $(\tau_{\I})^a (\tau_{\I})_a = -1$, we have
\begin{equation}
\label{eq:normalization_I}
- \kdd{t_{\I}}{\tau_{\I}}^2 + t_{\I}^{2p_3}\kdd{z_{\I}}{\tau_{\I}}^2 = -1 .
\end{equation}
The unit vector orthogonal to $\del \iv M_{\I}$, $(n_{\I})^a$, is obtained from 
the orthogonality with $(\tau_{\I})^a$, $(n_{\I})^a (\tau_{\I})_a = 0$, and 
the normalization, $(n_{\I})^a (n_{\I})_a =1$, such that
\begin{equation}
(n_{\I})^a 
= t_{\I}^{p_3} \dd{z_{\I}}{\tau_{\I}} \kvec{t_{\I}}^a 
+ t_{\I}^{-p_3} \dd{t_{\I}}{\tau_{\I}} \kvec{z_{\I}}^a 
\end{equation}
up to the overall signature. 
We can write the induced metric on $\del \iv M_{\I}$ as 
\begin{equation}
(q_{\I})_{ab} 
= - (\tau_{\I})_a (\tau_{\I})_b 
+ (\theta_{\I}^1)_a(\theta_{\I}^1)_b 
+  (\theta_{\I}^2)_a(\theta_{\I}^2)_b ,
\end{equation}
where $(\theta_{\I}^1)_a = t_{\I}^{p_1} (dx_{\I})_a$ and 
$(\theta_{\I}^2)_a = t_{\I}^{p_2} (dy_{\I})_a$. 
Then the orthonormal basis $\{ (\tau_{\I})^a, (e_{\I p})^a \}$ 
can be obtained by taking its dual. 
The extrinsic curvature on $\del \iv M_{\I}$,  defined by 
\eref{eq:def_extrinsic_curv}, can be obtained by a straightforward 
calculation: 
\begin{eqnarray}
\label{eq:extrinsic_curv_I}
(K_{\I})_{ab} 
&= t_{\I}^{p_3} \left\{ 
\dd{z_{\I}}{\tau_{\I}} \dd{^2 t_{\I}}{\tau_{\I}^2} 
- \dd{t_{\I}}{\tau_{\I}} \dd{^2 z_{\I}}{\tau_{\I}^2} 
- p_3 t_{\I}^{-1} \dd{z_{\I}}{\tau_{\I}} 
\left[ \kdd{t_{\I}}{\tau_{\I}}^2 + 1 \right] \right\}
(\tau_{\I})_a (\tau_{\I})_b  \nonumber\\ 
&\quad\, + \left( p_1 t_{\I}^{p_3 -1} \dd{z_{\I}}{\tau_{\I}} \right) 
(\theta_{\I}^1)_a (\theta_{\I}^1)_b  
+ \left( p_2 t_{\I}^{p_3 -1} \dd{z_{\I}}{\tau_{\I}} \right) 
(\theta_{\I}^2)_a (\theta_{\I}^2)_b .
\end{eqnarray}

%%%%%%%%%%%%%%%%  Subsubsection 4.1.2
\subsubsection{Compact quotient of vacuum Bianchi type VI$_0$ spacetime}

Bianchi VI$_0$ group is characterized by the Lie algebra
\begin{equation}
\fl 
[ (\xi_{\VI})_1 , (\xi_{\VI})_2 ] = 0 ,\quad 
[ (\xi_{\VI})_2 , (\xi_{\VI})_3 ] = -(\xi_{\VI})_1 ,\quad
[ (\xi_{\VI})_3 , (\xi_{\VI})_1 ] = (\xi_{\VI})_2
\end{equation}
for three Killing vectors $(\xi_{\VI})_i$. In terms of the coordinate 
basis $(\xi_{\VI})_i$'s are given by~\cite{RS75}
\begin{equation}
\fl 
(\xi_{\VI})_1 = \vect{x_{\VI}} ,\quad 
(\xi_{\VI})_2 = \vect{y_{\VI}} ,\quad 
(\xi_{\VI})_3 = \vect{z_{\VI}} - x_{\VI} \vect{x_{\VI}} + y_{\VI} \vect{y_{\VI}}  .
\end{equation}
The finite actions generated by these $(\xi_{\VI})_i$'s are given by
\begin{equation}
\label{eq:multiplication_VI}
\columnt{a}{b}{c} \columnt{x}{y}{z} 
= \columnt{a+ e^{-c}x}{b+ e^c y}{c+z} ,
\end{equation}
where the component vectors 
\begin{equation}
G_{\VI} = 
\left\{ \columnt{a}{b}{c} 
:= e^{c (\xi_{\VI})_3} e^{b (\xi_{\VI})_2} e^{a (\xi_{\VI})_1} \ |\ 
a,b,c \in \mathbb{R} \right\}
\end{equation}
form the Bianchi VI$_0$ group. 
The invariant dual basis of Bianchi VI$_0$ group is given by 
\begin{equation}
(\chi_{\VI})^1 = e^{z_{\VI}} dx_{\VI} ,\quad 
(\chi_{\VI})^2 = e^{-z_{\VI}} dy_{\VI} ,\quad 
(\chi_{\VI})^3 = dz_{\VI} .
\end{equation}

Then the invariant metric on $\Mh{\VI}$ inherited from its universal cover 
$\tMh{\VI}$ can be written as 
\begin{equation}
\label{eq:spatialmetric_VI}
(h_{\VI})_{ab} = (h_{\VI})_{ij} (\chi_{\VI}^i)_a (\chi_{\VI}^j)_b ,
\end{equation}
where $(h_{\VI})_{ij}$ is a constant matrix. 

Let $\tMg{\VI}$ be a vacuum, spatially homogeneous spacetime 
which has the Bianchi VI$_0$ group acting transitively on $\tM_{\VI}$. 
The general Bianchi VI$_0$ vacuum solution found 
by Ellis and MacCallum~\cite{EM69} is given by
\begin{eqnarray}
\label{eq:Ellis_sol}
(\tg_{\VI})_{ab} 
&= t_{\VI}^{-1/2} e^{Q^2 t_{\VI}^2} 
\left[ -(dt_{\VI})_a (dt_{\VI})_b + Q^{-2} (dz_{\VI})_a (dz_{\VI})_b \right] 
\nonumber\\ 
&\quad\, 
+ t_{\VI} \left[ e^{2z_{\VI}} (dx_{\VI})_a (dx_{\VI})_b 
+ e^{-2z_{\VI}} (dy_{\VI})_a (dy_{\VI})_b \right] ,
\end{eqnarray}
where $Q > 0$ is a constant.   

Now we consider the initial identification of $\tMg{\VI}$ which 
yields a SCLHS, $\Mg{\VI}$. For a compact quotient of $\tMh{\VI}$, 
we choose a compact manifold modeled on Sol whose fundamental 
group is given by
\begin{equation}
\pi_1(M_{\VI}) 
= \bk{ \alpha_{\VI},\beta_{\VI},\gamma_{\VI} }
{ [\alpha,\beta] = 1,\gamma\alpha\gamma\inv = \beta ,
\gamma\beta\gamma\inv = \beta^n \alpha\inv } ,
\end{equation} 
where $|n| > 2$. 
(This compact manifold is called ``f1/1(n)" in~\cite{KTH94}.)  
Its representation is obtained by taking conjugations with respect to 
the Bianchi VI$_0$ group which is 
$\Esom{\tM_{\VI}} = \Isom{\tM_{\VI}} = 
({\rm Sol\ and\ three\ extra\ discrete\ elements})$. 
For future use, we present one of the three discrete elements here. It is 
\begin{equation}
h : (x,y,z) \to (-x,-y,z) .
\end{equation}
Then we find that the representation are, up to conjugations, given by
\begin{equation}
\Gamma_{\VI} 
= \{ a_{\VI},b_{\VI},c_{\VI} \} 
= \left\{ \columnt{\alpha_0 u_1}{\alpha_0 u_2}{0} ,
\columnt{\alpha_0 v_1}{\alpha_0 v_2}{0} ,\columnt{0}{0}{c_3} \right\} ,
\end{equation}
for $n > 2$, and
\begin{equation}
\Gamma_{\VI} 
= \{ a_{\VI},b_{\VI},c_{\VI} \} 
= \left\{ \columnt{\alpha_0 u_1}{\alpha_0 u_2}{0} ,
\columnt{\alpha_0 v_1}{\alpha_0 v_2}{0} ,h \circ \columnt{0}{0}{c_3} \right\} ,
\end{equation}
for $n < -2$, with $\alpha_0 \in \mathbb{R}$. In these representations, 
$(u_1,v_1)$, $(u_2,v_2)$, and $c_3$ are determined in such a way 
that $\sign(n) e^{-c_3}$ and $\sign(n) e^{c_3}$ are the eigenvalues 
of matrix $\matrixs{0}{1}{-1}{n}$, and the corresponding 
normalized eigenvectors are $(u_1,v_1)$ and $(u_2,v_2)$, 
respectively. In particular, $e^{c_3} = | n + \sqrt{n^2 -4} |/2$. 

We cut out the spacetime $\Mg{\VI}$ along a timelike hypersurface 
$\Mq{\VI}$ whose spatial section, $\Ss{\VI}$, is taken to be 
the Killing orbits of $(\xi_{\VI})_1$ and $(\xi_{\VI})_2$. 
In this case, we should choose the generators $\alpha_I$ and 
$\beta_I$ as the generators of the boundary torus $\Sigma_{\VI} =T^2$. 
Thus, we obtain the representation on $\Sigma_{\VI}$,
\begin{equation}
\label{eq:initial_VI}
\Gamma_{\VI}|_{\Sigma_{\VI}} 
= \left\{\columns{\alpha_0 u_1}{\alpha_0 u_2} , 
\columns{\alpha_0 v_1}{\alpha_0 v_2} \right\} .
\end{equation}

Next we consider the local geometry of $\Mq{\VI}$. 
Since $(\xi_{\VI})_1 = \del/\del x_{\VI}$ and 
$(\xi_{\VI})_2 = \del/\del y_{\VI}$ lie in $\del \iv M_{\VI}$, 
the unit future-directed timelike vector $(\tau_{\VI})^a$ can be written as
\begin{equation}
(\tau_{\VI})^a 
= \kvec{\tau_{\VI}}^a 
= \dd{t_{\VI}}{\tau_{\VI}} \kvec{t_{\VI}}^a + \dd{z_{\VI}}{\tau_{\VI}} \kvec{z_{\VI}}^a .
\end{equation}
From $(\tau_{\VI})^a (\tau_{\VI})_a = -1$, we have
\begin{equation}
\label{eq:normalization_VI}
- t_{\VI}^{-1/2} e^{Q^2 t_{\VI}^2} 
\left[ \kdd{t_{\VI}}{\tau_{\VI}}^2 - Q^{-2} \kdd{z_{\VI}}{\tau_{\VI}}^2 \right] = -1 .
\end{equation}
The unit normal to $\del \iv M_{\VI}$, $(n_{\VI})^a$, is obtained from 
the orthogonality with $(\tau_{\VI})^a$, $(n_{\VI})^a (\tau_{\VI})_a = 0$, and 
the normalization, $(n_{\VI})^a (n_{\VI})_a =1$, such that
\begin{equation}
(n_{\VI})^a 
= Q^{-1} \dd{z_{\VI}}{\tau_{\VI}} \kvec{t_{\VI}}^a 
+ Q \dd{t_{\VI}}{\tau_{\VI}} \kvec{z_{\VI}}^a 
\end{equation}
up to the overall signature. 
We can write the induced metric on $\del \iv M_{\VI}$ as 
\begin{equation}
(q_{\VI})_{ab} 
= - (\tau_{\VI})_a (\tau_{\VI})_b 
+ (\theta_{\VI}^1)_a(\theta_{\VI}^1)_b + (\theta_{\VI}^2)_a(\theta_{\VI}^2)_b ,
\end{equation}
where $(\theta_{\VI}^1)_a = t_{\VI}^{1/2} e^{z_{\VI}} (dx_{\VI})_a$ 
and $(\theta_{\VI}^2)_a = t_{\VI}^{1/2} e^{-z_{\VI}} (dy_{\VI})_a$. 
Then the orthonormal basis $\{ (\tau_{\VI})^a, (e_{\VI p})^a \}$ 
can be obtained by taking its dual. 
The extrinsic curvature on $\del \iv M_{\VI}$, defined by 
\eref{eq:def_extrinsic_curv}, can be obtained as  
\begin{eqnarray}
\label{eq:extrinsic_curv_VI}
\fl 
(K_{\VI})_{ab} 
= \left[ t_{\VI}^{-1/2} e^{Q^2 t_{\VI}^2} Q^{-1} 
\left( \dd{z_{\VI}}{\tau_{\VI}} \dd{^2 t_{\VI}}{\tau_{\VI}^2} 
- \dd{t_{\VI}}{\tau_{\VI}} \dd{^2 z_{\VI}}{\tau_{\VI}^2}\right) 
- \frac12 (Qt_{\VI})^{-1} + 2 Q t_{\VI}  \right] 
(\tau_{\VI})_a (\tau_{\VI})_b  \nonumber\\
+ \left( \frac12 (Q t_{\VI})^{-1} \dd{z_{\VI}}{\tau_{\VI}} 
+ Q \dd{t_{\VI}}{\tau_{\VI}} \right) 
(\theta_{\VI}^1)_a (\theta_{\VI}^1)_b  \nonumber\\ 
+ \left( \frac12 (Q t_{\VI})^{-1} \dd{z_{\VI}}{\tau_{\VI}} 
- Q \dd{t_{\VI}}{\tau_{\VI}} \right) 
(\theta_{\VI}^2)_a (\theta_{\VI}^2)_b .
\end{eqnarray}

%%%%%%%%%%%%%%%%  Subsection 4.2
\subsection{Dynamics of the composite spacetime}
\label{subsec:dynamics}

We are in a position to proceed to glue compact quotients 
of vacuum Bianchi I and VI$_0$ given in the preceding subsections. 
We shall firstly consider the topological gluing, 
$\del \iv M_{\I} = \del \iv M_{\VI} = \del \iv M$. 
The initial identification can be represented, up to a finite rotation of 
the Killing vectors on $\Sigma_{\I} = \Sigma_{\VI} =: \Sigma$, as
\begin{equation}
\fl 
\Gamma|_{\Sigma_{\I}} = \Gamma|_{\Sigma_{\VI}} ,
\quad \iff \quad 
\columns{a_{\I}^1}{a_{\I}^2} = \columns{\alpha_0 u_1}{\alpha_0 u_2} 
,\quad
\columns{b_{\I}^1}{b_{\I}^2} = \columns{\alpha_0 v_1}{\alpha_0 v_2},
\end{equation}
from \eref{eq:initial_I} and \eref{eq:initial_VI}. 
Since $a_{\I}$ and $b_{\I}$ have not been fixed yet, we have
\begin{equation}
\label{eq:initial_glued}
\Gamma|_{\Sigma} 
= \left\{\columns{\alpha_0 u_1}{\alpha_0 u_2} , 
\columns{\alpha_0 v_1}{\alpha_0 v_2} \right\} .
\end{equation}

Next we consider the metric continuity on $\del \iv M$, 
$(q_{\I})_{ab} = (q_{\VI})_{ab}$. \eref{eq:metric_conti} can be represented, 
up to degrees of freedom of the choice of the orthonormal bases, as 
\begin{equation}
\fl 
(\tau_{\I})_a = (\tau_{\VI})_a ,\quad 
t_{\I}^{p_1} (dx_{\I})_a = t_{\VI}^{1/2} e^{z_{\VI}} (dx_{\VI})_a , \quad 
t_{\I}^{p_2} (dy_{\I})_a = t_{\VI}^{1/2} e^{-z_{\VI}} (dy_{\VI})_a .
\end{equation}
Although these equations have ambiguities of the homothetic 
transformations of coordinates, we interpret that 
such degrees of freedom have already been taken into account, i.e. 
$\tau_{\I} = \tau_{\VI} \equiv \tau$, $dx_{\I} = dx_{\VI}$, and 
$dy_{\I} = dy_{\VI}$. We have
\begin{equation}
t_{\I}^{p_1} = t_{\VI}^{1/2} e^{z_{\VI}}  , \quad
t_{\I}^{p_2} = t_{\VI}^{1/2} e^{-z_{\VI}} .
\end{equation}
From these, we obtain
\begin{equation}
t_{\VI} = t_{\I}^{p_1 + p_2}  , \quad 
z_{\VI} 
= \frac12 \ln t_{\I}^{p_1 -p_2} = \frac12 \ln t_{\VI}^P ,
\end{equation}
where we set $P = (p_1 - p_2)/(p_1 + p_2)$. 
Hence \eref{eq:normalization_VI} can be represented as 
\begin{equation*}
- t_{\VI}^{-1/2} e^{Q^2 t_{\VI}^2}
\left[ \Big( \dd{t_{\VI}}{\tau} \Big)^2 
- Q^{-2}\Big( \frac12 \dd{}{\tau} \ln t_{\VI}^P \Big)^2 \right] 
= -1 ,
\end{equation*}
thus, 
\begin{equation}
\label{eq:determine_orbit_IVI}
\kdd{t_{\VI}}{\tau}^2 
= \frac{ t_{\VI}^{1/2} e^{-Q^2 t_{\VI}^2 }}{ 1 - \frac{P^2}{4Q^2} t_{\VI}^{-2} } . 
\end{equation}
It is convenient to introduce the following parameterization for 
the Kasner parameters:
\begin{equation}
p_1 = \frac{1+u}{1+u+u^2} ,\quad 
p_2 = \frac{-u}{1+u+u^2} ,\quad
p_3 = \frac{u(1+u)}{1+u+u^2} ,
\end{equation}
where $u \in \mathbb{R}$. Using this parameterization, 
\eref{eq:determine_orbit_IVI} can be written as
\begin{equation}
\label{eq:det_orbit_IVI}
\kdd{t_{\VI}}{\tau}^2 
= \frac{ t_{\VI}^{1/2} e^{-Q^2 t_{\VI}^2 }}{ 1 - \frac{u^2}{Q^2} t_{\VI}^{-2} } .
\end{equation}
The sufficient condition of the existence of solutions 
of \eref{eq:det_orbit_IVI} is that: 
\begin{equation}
t_{\VI} > \frac{|u|}{Q} .
\end{equation}
We assume that $dt_{\VI}/d\tau > 0$, i.e. $t_{\VI}$ increases 
in the same direction with $\tau$. 
Since \eref{eq:det_orbit_IVI} is the first order differential equation, 
there exists a unique solution. The solution yields an orbit of the shell 
$(t_{\VI}(\tau), z_{\VI}(\tau))$. 
We shall note that it follows from \eref{eq:det_orbit_IVI} that 
$dt_{\VI}/d\tau$ is decreasing. 
We will consider the consequence of this behavior below. 

Finally, we solve the junction condition 
\eref{eq:H_constraint}, \eref{eq:M_constraint}, \eref{eq:evolution}. 
The role of the junction condition here is that it determines 
what matter is compressed on $\del \iv M$
\footnote[1]{The role of the junction condition is different from 
that of a spherical shell. The dynamics of the toroidal shell have already been 
determined by \eref{eq:metric_conti}, 
while that of the spherical shell is determined by 
\eref{eq:H_constraint}, \eref{eq:M_constraint}, \eref{eq:evolution}.
This may be caused by the fact that no Teichm\"uller deformation 
is induced by the instantaneous move of the homogeneous toroidal shell in $T^3$.}. 
We can write the surface energy-momentum tensor $S_{ab}$ 
in the diagonalized form,
\begin{equation}
\label{eq:surface_em_tensor}
S_{ab} 
= \rho(\tau) \tau_a \tau_b 
+ P_1(\tau) (\theta^1)_a (\theta^1)_b + P_2(\tau) (\theta^2)_a (\theta^2)_b ,
\end{equation}
where $\rho$ is the surface energy density and $P_p$ ($p=1,2$) 
the principal pressure with respect to the orthonormal dual basis 
$(\theta^p)_a$ on $\Sigma$. 
Since $\Sigma$ is locally homogeneous, $\rho$ and $P_p$ depend only on $\tau$. 
These three functions is determined by the junction condition. 
In fact, we need only \eref{eq:evolution} with \eref{eq:surface_em_tensor}, 
\begin{equation}
\fl 
[K_{ab}] 
= -4 \pi \left[ (\rho + P_1 + P_2)(d\tau)_a (d\tau)_b 
+ (\rho + P_1 - P_2)(\theta^1)_a (\theta^1)_b 
+ (\rho - P_1 + P_2)(\theta^2)_a (\theta^2)_b \right]
\end{equation}
Thus, $\rho$, $P_1$ and $P_2$ is determined as 
\begin{eqnarray}
\rho &= \frac{1}{8\pi} \left( [K_{11}] + [K_{22}] \right) ,\\
P_1 &= \frac{1}{8\pi} \left( [K_{22}] - [K_{00}] \right) ,\\
P_1 &= \frac{1}{8\pi} \left( [K_{11}] - [K_{00}] \right) ,
\end{eqnarray}
where $00$-component denotes the $\tau\tau$-component. 
Applying \eref{eq:extrinsic_curv_I} and \eref{eq:extrinsic_curv_VI} to $[K_{ab}]$, 
we have following equations:
\begin{eqnarray}
\fl 
[K_{00}] = \left\{ t_{\I}^{p_3} \left( 
       \dd{z_{\I}}{\tau_{\I}} \dd{^2 t_{\I}}{\tau_{\I}^2} 
	   - \dd{t_{\I}}{\tau_{\I}} \dd{^2 z_{\I}}{\tau_{\I}^2} 
	   - p_3 t_{\I}^{-1} \dd{z_{\I}}{\tau_{\I}} 
	   \left[ \kdd{t_{\I}}{\tau_{\I}}^2 + 1 \right]
       \right) \right\} \nonumber\\ 
\fl \qquad\quad\  
      - \left\{ t_{\VI}^{-1/2} e^{Q^2 t_{\VI}^2} Q^{-1} 
      \left( \dd{z_{\VI}}{\tau_{\VI}} \dd{^2 t_{\VI}}{\tau_{\VI}^2} 
	  - \dd{t_{\VI}}{\tau_{\VI}} \dd{^2 z_{\VI}}{\tau_{\VI}^2}
	  \right) - \frac12 (Qt_{\VI})^{-1} + 2 Q t_{\VI}  
	  \right\} ,
\end{eqnarray}
\begin{eqnarray}
\fl
[K_{11}] = \left\{ p_1 t_{\I}^{p_3 -1} \dd{z_{\I}}{\tau_{\I}} \right\}
                  - \left\{ \frac12 (Q t_{\VI})^{-1} \dd{z_{\VI}}{\tau_{\VI}} 
                  + Q \dd{t_{\VI}}{\tau_{\VI}} \right\} ,\\ 
\fl 
[K_{22}] = \left\{ p_2 t_{\I}^{p_3 -1} \dd{z_{\I}}{\tau_{\I}} \right\} 
                  - \left\{ \frac12 (Q t_{\VI})^{-1} \dd{z_{\VI}}{\tau_{\VI}} 
               - Q \dd{t_{\VI}}{\tau_{\VI}} \right\} ,
\end{eqnarray}
up to the overall signatures of each $(K_{\I})_{ab}$ and 
$(K_{\VI})_{ab}$, i.e. we may change the signature in front of each bracket. 
The $\tau$-derivatives in the above equations have 
been determined by \eref{eq:det_orbit_IVI} 
(and \eref{eq:normalization_I}), and they should be read as given 
functions of $\tau$. 

To discuss whether the composite spacetime consists of realistic matter, 
we shall see the energy condition by writing down the energy density on the shell 
$\del \iv M$. We have
\begin{equation}
\label{eq:energy_density}
\fl 
\rho 
= \frac{1}{8\pi} \left(\pm \frac{1}{t_{\VI}^{1+u+u^2}} 
\sqrt{ t_{\VI}^{2u(1+u)} \kdd{t_{\VI}}{\tau}^2 -\frac{1}{(1+u+u^2)^2} } 
- \frac{1}{2Q} \frac{1}{t_{\VI}^2} \dd{t_{\VI}}{\tau} \right) ,
\end{equation}
where \eref{eq:det_orbit_IVI} is satisfied, and hence 
$t_{\VI}$ is given implicitly as a function of the proper time $\tau$ on 
the shell, $t_{\VI} = t_{\VI}(\tau)$. Since there are the degrees of freedom of 
taking the signatures of the unit normal to $\del \iv M_{\I}$ and 
$\del \iv M_{\VI}$, and hence those of the extrinsic curvatures, 
we can control $\rho$ to be always positive. As \eref{eq:det_orbit_IVI} indicates 
that $dt_{\VI}/d\tau$ is monotonically decreasing, however, 
we find that the first term in the right hand side of \eref{eq:energy_density} 
will become complex after finite-time evolution. 
If we regard such a complex energy unphysical, 
it is understood that the shell $\del \iv M$ stops. 
Nevertheless, it is natural to consider each building block 
independently continues evolving after that. 
Thus, it is fairly to say that this phenomenon may occur due to the break down of 
the thin-shell approximation. 

Finally, up to the problem indicated above, 
we can find the Teichm\"uller deformation of the boundary torus. 
The result is 
\begin{eqnarray}
r = \frac{ \sqrt{(u_1)^2 t_{\VI}(\tau)^P + (u_2)^2 t_{\VI}(\tau)^{-P}} }
{ \alpha_0 | u_1 v_2 - v_1 u_2 | } , \\
s = 
\frac{ u_1 v_1 t_{\VI}(\tau)^P + u_2 v_2 t_{\VI}(\tau)^{-P} }
{ \alpha_0  | u_1 v_2 - v_1 u_2 | 
\sqrt{(u_1)^2 t_{\VI}(\tau)^P + (u_2)^2 t_{\VI}(\tau)^{-P}} } ,
\end{eqnarray}
where $r$ and $s$ are the Teichm\"uller parameters defined 
in \eref{eq:Teich_torus}. 
If $t_{\VI}(\tau)$ can become infinitely large, both deformation parameters 
diverge. 
Also in the case that $P = 0$, i.e., the Bianchi type I region becomes plane 
symmetric, no Teichm\"uller deformation occurs.

%%%%%%%%%%%%%%%%  Section 5
\section{Summary and Discussions}
\label{sec:smry_dcs}

Motivated by the Geometrization Conjecture for 3-manifolds, 
we give a new formulation for the construction of 
spatially compact composite spacetimes. 
It is the direct application of the construction of the SCLHSs formulated by 
TKH~\cite{TKH97a} and the junction method of Israel~\cite{ISRAEL66}. 
The key element for it is that, to keep the local homogeneity of the building blocks 
(SCLHSs) we assume that inhomogeneities which should arise by the gluing of 
two different geometries smoothly is compressed onto the matter 
on the timelike shell. The priority of this method is that 
geometrical part of the dynamical degrees of freedom are kept finite so that 
we can explicitly construct a composite spacetime as a solution of 
the Einstein equations. 

From the examination of the behavior of a composite spacetime 
constructed in such a way by gluing compact quotients of 
vacuum Bianchi type I and VI$_0$ together along the timelike shell, however, 
we find the strange behavior of the timelike shell. 
Indeed, it terminates its evolution in finite time whereas 
the building blocks would continue evolving provided that 
we require the realistic matter on the shell. 
We interpret this such that our approximation, the thin-shell approximation, 
may not be so flexible as to describe the gluing sufficiently. 

Also, the orbit of the shell of our example is determined 
in a different way from the spherically symmetric junctions. 
In the latter cases, the surface stress-energy tensor should be assumed 
and the junction conditions has a role for determining the orbit of 
the shell~\cite{ISRAEL66,SSHELLS}. 
We think that the difference occurs because $T^3$ topology which 
we assume for the topology of the compact quotient of Bianchi type I 
does not have any curvature scale. 
We can speculate that when we consider the gluing of compact quotients of 
Nil and Sol, the dynamics will be determined in a similar way with 
the spherically symmetric junctions.  

The gluing procedure we give in this article is very restrictive 
so that the resulting spacetime must contain homogeneous tori. 
In fact, only from such a single gluing, the spacetime admits 
a geometric structure, i.e. it should be interpreted as 
the inhomogeneization of the SCLHS as in the case of the Gowdy spacetime. 
Thus, our model cannot change its topology without the combinations with other gluing. 
However, we remark that there is a non-trivial geometrical meaning for 
the resulting compact 3-manifolds. 
They are the \emph{graph manifolds} which, roughly speaking, consist of 
the Seifert fibered geometries and Sol glued together 
along tori~\cite{ANDERSON97}. 
Since we have assumed the homogeneity of the boundary torus 
as a technical assumption to deal with the Einstein equations analytically, 
it should be removed to get the non-trivial torus sum. 

To end this article, we comment about the smooth gluing to obtain 
a spacetime with generic topology. 
In the locally spherically symmetric case, Morrow-Jones and 
Witt~\cite{MJW93} have developed a method for getting a smooth gluing 
by taking into account a local deformation of the geometry 
in the direction orthogonal to the surface of symmetry. 
(Such a local deformation may also be interpreted as an example of 
the spacetimes containing a thick shell as mentioned in 
Section~\ref{sec:formalism}. ) It is expected that 
by considering models which admit only two-dimensional homogeneity, 
e.g. the Gowdy models~\cite{GOWDY74,TANIMOTO98}, 
it might be possible to consider a smooth gluing of the torus boundary 
in the same way. 

\ack 
The authors are very grateful to M.~Tanimoto for helpful discussions and 
for detailed comments. 
They also thank A.~Hosoya and H.~Ishihara for continuous encouragement. 
K.Y. acknowledges financial support from 
the Japan Society for the Promotion of Science 
and the Ministry of Education, Science, Sports and Culture.


\Bibliography{99}
\bibitem{GOWDY74} 
Gowdy R H 1974 \APNY {\bf 83} 203
\bibitem{TANIMOTO98} 
Tanimoto M 1998 \JMP {\bf 39} 489
\bibitem{TANIMOTO00}
Tanimoto M 2000 {\it Preprint} gr-qc/0003033
\bibitem{THURSTON82} 
Thurston W P 1982 {\it Bull. Amer. Math. Soc.} {\bf 6} 357 
\bibitem{SCOTT83} 
Scott P 1983 {\it Bull. London Math. Soc.} {\bf 15} 401 
\bibitem{MJW93} 
Morrow-Jones J and Witt D M 1993 \PR D {\bf 48} 2516
\bibitem{RS75} 
Ryan M P and Shepley L C 1975 
{\it Homogeneous Relativistic Cosmologies} 
(Princeton: Princeton University Press)
\bibitem{AS91} 
Ashtekar A and Samuel J 1991 \CQG {\bf 8} 2191
\bibitem{FIK93} 
Fujiwara Y, Ishihara H and Kodama H 1993 \CQG {\bf 10} 859
\bibitem{KTH94} 
Koike T, Tanimoto M and Hosoya A 1994 \JMP {\bf 35} 4855
\bibitem{TKH97a} 
Tanimoto M, Koike T and Hosoya A 1997 \JMP {\bf 38} 350
\bibitem{TKH97b} 
Tanimoto M, Koike T and Hosoya A 1997 \JMP {\bf 38} 6550
\bibitem{KODAMA98} 
Kodama H 1998 {\it Prog. Theor. Phys.} {\bf 99} 173
\bibitem{ISRAEL66} 
Israel W 1966 {\it IL Nuovo Cimento} {\bf 44B} 1
\bibitem{EM69} 
Ellis G F R and MacCallum M A H 1969 {\it Commun. Math. Phys.} 
{\bf 12} 108
\bibitem{SK94}
Sato H and Kodama H 1994 {\it General Relativity} 
(Tokyo: Iwanami Shoten) 
\bibitem{WALD84} 
Wald R M 1984 {\it General Relativity} 
(Chicago: Chicago University Press) 
\bibitem{SSHELLS}
Sato H 1986 {\it Prog. Theor. Phys.} {\bf 76} 1250 \\
Maeda K 1986 {\it Gen. Rel. Grav.} {\bf 18} 931 \\
Blau S K, Guendelman E I and Guth A H 1987 \PR D {\bf 35} 1747 \\
Berezin V A, Kuzmin V A and Tkachev I I 1987 \PR D {\bf 36} 2919 \\
Sakai N and Maeda K 1994 \PR D {\bf 50} 5425 \\
\bibitem{THURSTON97} 
Thurston W P {\it Three-Dimensional Geometry and Topology} vol 1, 
ed S Levy (Princeton, NJ: Princeton University Press)
\bibitem{ANDERSON97} 
Anderson M in {\it Comparison geometry, Math. Sci. Res. Inst. Publ.} 
vol 30 (Cambridge: Cambridge University Press) 49
\endbib
\end{document}